# Effect of changing the rare earth cation type on the structure and crystallization behavior of an aluminoborosilicate glass

A. Quintas, D. Caurant, O. Majérus
*Laboratoire de Chimie de la Matière Condensée de Paris, UMR 7574, ENSCP, Paris, France*

A. Quintas, J-L. Dussossoy
*Commissariat à l'Energie Atomique, Centre d'étude de la Vallée du Rhône,
DEN/DTCD/SCDV/LEBV, Bagnols-sur-Cèze, France*

T. Charpentier
*CEA Saclay, Laboratoire de Structure et Dynamique par Résonance Magnétique,
DSM/DRECAM/SCM – CEA CNRS URA 331, Gif-sur-Yvette, France*

An aluminoborosilicate glass, containing high amount of rare earth (RE) accordingly to the following composition 50.68 $SiO_2$ – 4.25 $Al_2O_3$ - 8.50 $B_2O_3$ – 12.19 $Na_2O$ – 4.84 $CaO$ – 3.19 $ZrO_2$ – 16.35 $RE_2O_3$ (wt.%), is currently under study for the immobilization of nuclear waste solutions. In this work, we wanted to investigate the effect of changing the RE cation type on the glass structure and on its crystallization behavior. For this purpose, a glass series was elaborated in which the nature of the RE is varying from lanthanum to lutetium. In this glass series, only little effect was observed on the glass structure. On the contrary, a strong impact was put in evidence on the crystallization behavior through different heat treatments. A slow cooling of the melt at 1°C/min, revealed significant crystallization of apatite $Ca_2RE_8(SiO_4)_6O_2$ in sample containing rare earths with ionic radii close to that of calcium. Another heat treatment consisting of successive nucleation and growth stages, performed to force the crystallization in the bulk and reduce any surface crystallization effect, put in evidence the existence of a strongly heterogeneous second rare earth rich silicate phase for samples containing RE with low ionic radius (from Y to Lu).

## Introduction

In order to immobilize radioactive wastes, stemming from the reprocessing of high discharge burn up nuclear fuel, a new confinement glass matrix has been recently envisaged. A rare-earth (RE) rich glass has been selected for its good performances (melting at 1300°C, good chemical durability and waste capacity, low crystallization tendency) [1,2]. A simplified seven oxides glass model composition (referred to as glass A): 61.81 $SiO_2$ - 3.05 $Al_2O_3$ - 8.94 $B_2O_3$ - 14.41 $Na_2O$ - 6.33 $CaO$ - 1.90 $ZrO_2$ - 3.56 $RE_2O_3$ (mol.%) was adopted to facilitate crystallization studies and structural investigations. In previous studies [3-5] of glass A, only Nd or La were used as rare earth in the glass composition. Heat treatments performed on both rare earth glasses revealed significant differences between Nd and La rich glasses in the crystallization of an apatite phase (a much lower crystallization rate was obtained with the La bearing glasses) which was the first demonstration of the non-equivalent role of Nd and La towards the crystallization tendency.

In this work, we investigate how the nature of the rare earth may influence the glass structure and its crystallization behavior. For this purpose, an extended series of ten glasses has been prepared in which the nature of the trivalent RE is varied from lanthanum to lutetium. The following lanthanides, classified in the order of decreasing ionic radius and increasing field strength, were independently incorporated in the glass composition: La, Pr, Nd, Sm, Eu, Gd, Er, Yb, Lu. In addition, a glass containing diamagnetic yttrium, which is a trivalent rare earth that does not belong to the previous lanthanide series, was also prepared (in terms of ionic radius, $Y^{3+}$ inserts between $Gd^{3+}$ and $Er^{3+}$). All of these glasses (except the Nd bearing glasses) contain 0.15 mol.% of $Nd_2O_3$ introduced as a local structural probe. The influence of the RE cation type on the glass structure was studied by the mean of Raman and MAS NMR.

## Results and discussion

Both MAS NMR and Raman spectroscopies did not reveal significant effect of a change of the rare earth cation type on the glass structure. Only a slight effect on the boron speciation was noticed. According to $^{11}B$ MAS NMR (figure 1) spectroscopy, the $[BO_4]/[BO_3]$ ratio decreases with increasing rare earth cationic field strength (the proportion of fourfold boron species among total boron species varies from 41.6% for La to 32.9% for Lu).

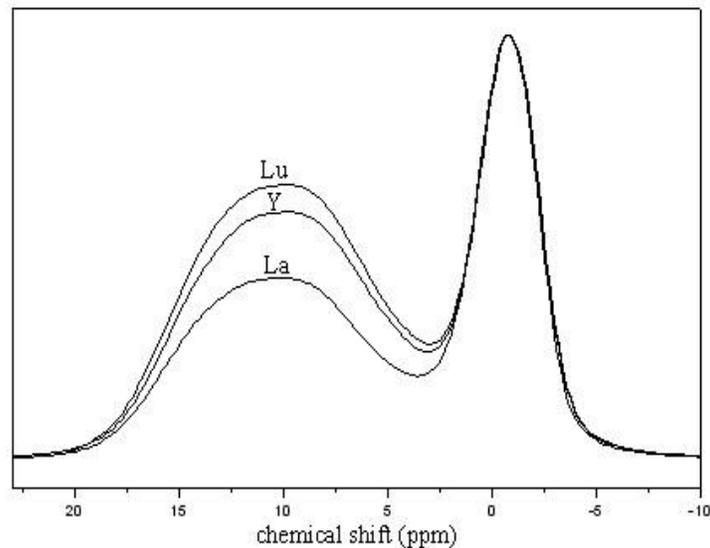

Figure 1 : $^{11}B$ MAS NMR spectra recorded for the diamagnetic RE (La, Y and Lu) and normalized to the same maximum $BO_4$ intensity ($B_0$=11.75T, $\nu_{rot}$=12.5kHz).

In the meantime, crystallization studies were intended to follow the impact of the nature of the RE on the crystallization behavior of the glass melt. Indeed, a high crystallization tendency during the cooling of the melt is considered as a drawback for the specific application envisaged here. The incorporation of α-radionuclides in crystals may induce amorphization of the crystalline phases and the resulting swelling could lead to waste form fracturation that can affect the long term durability. Two different heat treatments were performed to study crystallization:

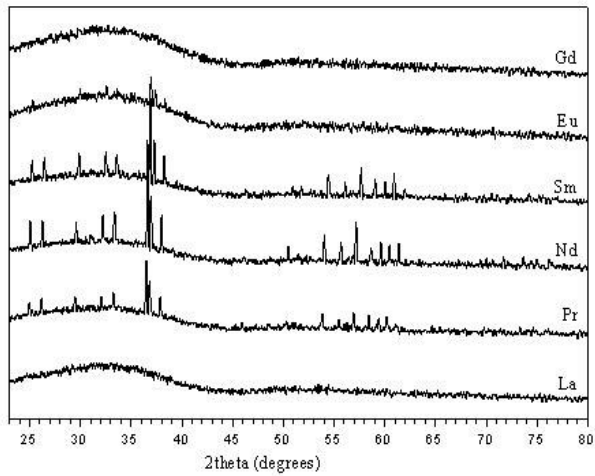 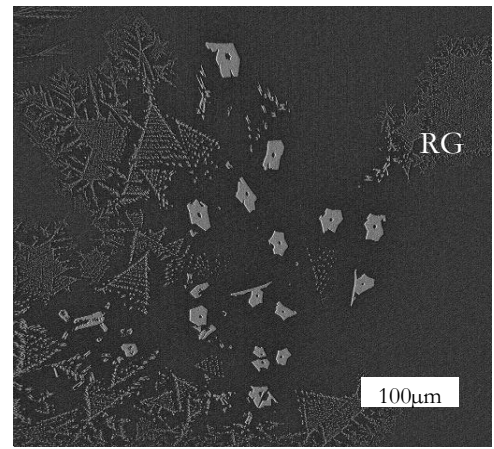

Figure 2 : X-ray diffraction patterns of the RE glass series restricted from La to Gd after slow cooling of the melt at 1°C/min ($\lambda_{CoK\alpha 1}$=1.789Å).

Figure 3 : Back-scattered SEM image of the neodymium bearing glass showing silicate apatite crystals with different microstructures formed during slow cooling of the melt (1°C/min); RG: residual glass.

- A slow cooling of the melt (1°C/min from 1350°C to room temperature) was done in order to simulate the natural cooling in the bulk of nuclear waste glass containers after casting and to determine the nature and the amount of the crystalline phases that may form during slow cooling. It appears that crystallization is strongly affected by the nature of the RE present in the glass composition. According to the diffractogramms presented in figure 2, glasses containing RE with smaller ionic radius than that of Eu do not lead to crystallization of apatite $Ca_2RE_8(SiO_4)_6O_2$ phase (figure 3). La-bearing glass also displays very low crystallization tendency.

- Another heat treatment consisting of successive nucleation (2 hours at Tg+20°C) and growth (30 hours at 934°C) stages was intended to force the crystallization in the bulk and to reduce any surface crystallization effect. Compared to the slow cooling, this heat treatment gives way to the crystallization of apatite crystals from La to Gd. Note that the progressive shift of the diffraction peaks towards higher angle from La to Gd (i.e. increase of the cell dimensions) is consistent with the decrease of the ionic radius of the rare earth. As for the slow cooling heat treatment, a maximum crystallization rate is obtained for glasses containing Pr or Nd. Moreover, another crystalline phase (not detected by XRD on massive sample) can be observed from yttrium to lutetium and seems to grow at the expense of the apatite phase. The nucleation process of this so-called P-phase is strongly heterogeneous as it only appears on a very thin surface layer. The extent of crystallization of the P-phase was enhanced by performing the nucleation and growth heat treatment in the same conditions on powder glass samples (particle size: 80-125µm). The diffractogramms of the powder samples exhibit peaks of the P-phase already observed in previous studies in glasses derived from glass A and containing high amount of calcium substituting for sodium. This crystalline phase was then identified as a calcium and neodymium silicate of composition $Ca_{1-3x}Nd_{2x}SiO_3$ (where x≈0.18) determined by electron probe microanalysis.

In order to account for the maximum crystallization observed for Pr and Nd glass samples, we suggest a thermodynamic approach based on the comparison of ionic radius between $RE^{3+}$ and $Ca^{2+}$. In the apatite structure, $RE^{3+}$ and $Ca^{2+}$ may occupy the two 4f and 6h sites (respectively 9-fold and 7-fold coordinated). During the nucleation stage, the apatite nuclei form at relatively low temperature (Tg+20°C). So the diffusion process is slow and as the initial glass system is disordered, it is assumed that the nuclei formed during the first stage of nucleation are disordered (random distribution of $RE^{3+}$ and $Ca^{2+}$ cations in both 4f and 6h sites). In the slow cooling heat treatment, the nuclei form at quite high temperature. In this case, we suppose that the first nuclei are also disordered in order to lower the entropy difference between liquid and nucleus (-TΔS). The enthalpy of the first disordered nucleus is lower in the case of similar ionic radii (fewer local lattice distortion) than in the case of dissimilar ionic radii. Subsequently, at a given temperature, when the RE ionic radius is close to that of calcium, the difference in free energy between the liquid and the crystal is greater which lead to a greater crystallization driving force.

**Conclusion**

In conclusion, the glass network is only weakly affected by a change of the rare earth cation type. On the contrary, the crystallization behavior dramatically depends on the nature of the rare earth introduced in the glass composition. During slow cooling of the melt, the formation of an apatite crystalline phase is strongly favored when the ionic radius of the incorporated rare earth is close to that of $Ca^{2+}$ (Pr, Nd or Sm) which can be accounted for by thermodynamic arguments. An isotherm heat treatment enhances the global crystallization tendency. With this last heat treatment, a new surface crystalline phase appears for glasses with low apatite crystallization (glasses containing low RE ionic radius: from Y to Lu).